\documentclass[a4paper,11pt]{article}

\usepackage{jheppub} 
\usepackage{graphicx}
\usepackage{amsmath}
\usepackage[utf8]{inputenc}

\usepackage{amssymb}
\usepackage{amsfonts}
\usepackage{amsbsy}
\usepackage{url}
\usepackage{enumerate}
\usepackage{caption}
\usepackage{subcaption}
\usepackage{color}
\usepackage{bm}
\usepackage{bbm}
\usepackage{comment}
\usepackage{multirow,bigdelim}
\usepackage[head1]{DotArrow}
\usepackage[normalem]{ulem}
\usepackage{CJKutf8}
\usepackage{braket}

\makeatletter
\@addtoreset{equation}{section}
\makeatother

\makeatletter
\gdef\@fpheader{}
\makeatother

\graphicspath{{./}{./figs/}}
\newbox{\ORCIDicon}
\title{
Comparison of the Migdal transition probabilities in electron-atom inelastic cross sections
}

\author[a]{Wakutaka Nakano}
\emailAdd{wnakano@post.kek.jp}
\affiliation[a]{KEK Theory Center, Tsukuba 305-0801, Japan}

\abstract{
The Migdal transition probabilities for dark matter scattering are compared to the total single-electron inelastic cross sections of electron-atom scattering for isolated Ar and Xe. The comparison is done by expressing the electron-atom scattering cross section by connecting the Migdal probability. The resultant differences are around $30 \ \%$ for Ar and $80 \ \%$ for Xe in $\sim$ 1 keV of incoming electron energy. The transition is dominated from the $3p$ shell electrons for Ar. For Xe, $5p$ states dominate the contribution, but, $4d$ states give $\sim 40\ \%$ contribution at 1 keV.
}

\preprint{KEK-TH-2659}

\begin{document}

\maketitle

\section{Introduction}

Dark matter is unknown physics beyond the standard model indicated from astrophysical and cosmological observations (see, e.g., Ref.~\cite{Murayama:2007ek} for review). Dark matter is often assumed to have interactions with the standard model particles. Then, dark matter can be detected by the interaction with probe atoms. The benchmark of dark matter model is called the weakly interacting massive particles (WIMPs) which interacts with nuclei~\cite{Goodman:1984dc} and the abundance is set by the thermal freeze out~\cite{Lee:1977ua}. Direct detection experiments search the WIMP signals from the detection of ionization, scintillation, and heat in the detectors (see Ref.~\cite{Lewin:1995rx} for review).

The Migdal effect~\cite{migdal1939ionizatsiya,migdal1941ionizacia,Landau:1991wop} is an inelastic process of scattered nucleus. It causes the excitation/ionization of electrons from the nuclear recoil or decay. The Migdal effect is observed in $\alpha$ decay~\cite{Rapaport:1975zza,Rapaport:1975zz,fischbeck1975spectroscopy,fischbeck1977angular} and $\beta$ decay~\cite{boehm1954internal,berlovich1965investigation,couratin2012first,Lienard:2015joa,Fabian:2018pin}, however, not observed in nuclear scattering. Now, experiments for detecting the Migdal effect from gas probes are proceeding~\cite{Nakamura:2020kex,Araujo:2022wjh}. 

The Migdal effect is also effective for searching small energy signals from dark matter. The Migdal effect occurs as an inelastic process of dark matter-nucleus scattering and causes the excitation/ionization of electrons from the nuclear recoil. The excitation/ionization of electrons returns electromagnetic signals that are easier to detect than heat in the case of elastic scattering. Recently, the Migdal effect is studied or used in many papers~\cite{baur1983ionisation,vegh1983multiple,wauters1997recoil,berakdar2001electronic,Vergados:2005dpd,Moustakidis:2005gx,Ejiri:2005aj,talman2006excitations,Bernabei:2007jz,liertzer2012multielectron,Vergados:2013raa,pindzola2014neutron,Sharma:2017fmo,Ibe:2017yqa,Dolan:2017xbu,LUX:2018akb,Bell:2019egg,GrillidiCortona:2020owp,vonKrosigk:2020udi,XMASS:2022tkr,Coskuner:2018are,EDELWEISS:2019vjv,Tvrznikova:2019tgx,CDEX:2019hzn,XENON:2019zpr,Baxter:2019pnz,Essig:2019xkx,Liang:2019nnx,Kozaczuk:2020uzb,SENSEI:2020dpa,Liu:2020pat,Nakamura:2020kex,Pindzola:2020whx,Kahn:2020fef,Knapen:2020aky,Knapen:2021bwg,Liang:2022xbu,LUX:2020yym,Liang:2020ryg,He:2020sat,Flambaum:2020xxo,Akerib:2021pfd,Collar:2021fcl,Bell:2021zkr,Liao:2021yog,Menshikov:2021jxa,Acevedo:2021kly,COSINE-100:2021poy,Berlin:2021zbv,CDEX:2021cll,Wang:2021oha,Bell:2021ihi,Abbamonte:2022rfh,Chatterjee:2022gbo,EDELWEISS:2022ktt,Angevaare:2022zhy,Campbell-Deem:2022fqm,Araujo:2022wjh,DarkSide:2022dhx,Blanco:2022pkt,Cox:2022ekg,GlobalArgonDarkMatter:2022ppc,Tomar:2022ofh,Adams:2022zvg,Berghaus:2022pbu,Li:2022acp,NEWS-G:2023qwh,DarkSide-50:2023fcw,SuperCDMS:2023sql,Bell:2023uvf,Tilly:2023fhw,AtzoriCorona:2023ais,Qiao:2023pbw,Xu:2023wev,LZ:2023poo,PandaX:2023xgl,Yun:2023huf,Gu:2023pfg,Li:2023xkf,Herrera:2023xun,Kim:2023lkb,Dzuba:2023zcq,SENSEI:2023zdf,Liang:2024tef,He:2024hkr,Balan:2024cmq,MIGDAL:2024alc,Kang:2024kec,Ibarra:2024mpq,Lee:2024wzd}, which includes that the comparison~\cite{LUX:2018akb,Bell:2019egg,GrillidiCortona:2020owp,vonKrosigk:2020udi,XMASS:2022tkr} to bremsstrahlung of photon~\cite{Kouvaris:2016afs,Millar:2018hkv}, theoretical comparison to the dark matter-electron scattering~\cite{Baxter:2019pnz,Essig:2019xkx}, the Migdal-photoabsorption relation~\cite{Liu:2020pat}, formulation for semiconductor response~\cite{Knapen:2020aky,Knapen:2021bwg,Liang:2022xbu}, and the semi-inclusive calculation~\cite{Cox:2022ekg}.

No detection of the Migdal effect from nuclear recoil in liquid Xe probes is reported from Ref.~\cite{Xu:2023wev}. The no detection may caused by the inaccurate predictions for the Migdal transition probability or the signal response in liquid Xe. In this paper, the Migdal probability is compared to the total single-electron  inelastic cross section of electron-atom scattering for isolated Ar and Xe. The comparison is possible by expressing the electron-atom scattering cross section by the Migdal probability calculated in Ref.~\cite{Ibe:2017yqa}. 

The energy eigenstates of atom are given in Sec.~\ref{sec:state_of_atom}. The formalism uses the center of mass coordinates and is useful for treating long-range interactions. The interaction potential is derived in Sec.~\ref{sec:pot}. The interactions are approximately described by the electron-electron interaction. Therefore, the shape of potential is the Coulomb one. The total excitation and ionization cross section is calculated in Sec.~\ref{sec:cross_section}. Especially, the cross section is expressed by the Migdal transition factor of dark matter-atom scattering for single electron excitation/ionization with the dipole approximation given in Ref.~\cite{Ibe:2017yqa}. Then, the result is compared with a database value which is a combination of experimental and theoretical data and other theoretical calculations. The natural unit is used in this paper.

\section{Energy eigenstates of atomic system}
\label{sec:state_of_atom}

The Hamiltonian of atom is 
\begin{align}
    H_{\rm atom} = \frac{\Vec{p}_{n}\,^2}{2m_n} + \sum_{i=1}^Z \frac{ \Vec{p}_{e,i}\,^2}{2m_e} -Z\alpha \sum_{i=1}^Z \frac{1}{|\vec{r}_{e,i}-\vec{r}_n|} + \alpha \sum_{\{i,j\}} \frac{1}{|\vec{r}_{e,i}-\vec{r}_{e,j}|} ,
\end{align}
where $\Vec{p}_{n}$ and $\Vec{p}_{e,i}$ is the three-dimensional momentum of the nucleus and $i$-th electron, $\vec{r}$ is the three-dimensional coordinates, $m_n$ and $m_e$ are the mass of the nucleus and electron, $Z$ is the atomic number, and $\alpha$ is the fine structure constant. The bracket $\{i,j\}$ represents the sum of the pair $i$ and $j$ under $i < j$.

The above Hamiltonian can be expressed by the center of mass coordinates of atom as
\begin{align}
        H_{\rm atom} \approx & \frac{\Vec{p}_{\rm CM}\,^2}{2m_A} + \sum_{i=1}^Z \frac{\Vec{p}_{i}\,^2}{2m_e} -Z\alpha \sum_{i=1}^Z \frac{1}{|\vec{r}_{i}|} + \alpha \sum_{\{i,j\}} \frac{1}{|\vec{r}_{i}-\vec{r}_{j}|} \\
        = & \frac{\Vec{p}_{\rm CM}\,^2}{2m_A} + H_{\rm re} 
\end{align}
with $m_e \ll m_n$. Here, $\Vec{r}_{\rm CM}=(m_n\Vec{r}_n+m_e\sum_i \Vec{r}_{e,i})/(m_n+Zm_e)$ and $\Vec{r}_i=\Vec{r}_{e,i}-\Vec{r}_{\rm CM}$ are the center of mass and relative coordinates, $\Vec{p}_{\rm CM} \approx m_A (\partial_t \Vec{r}_{\rm CM})$ and $\Vec{p}_i \approx m_e (\partial_t \Vec{r}_i)$ are their momentum, and $m_A \approx m_n+Zm_e$ is the atomic mass. 

The above Hamiltonian can be separated by the center of mass energy and the others. Therefore, the energy eigenstates of the atom are given by $|\Psi_A\rangle = |\Vec{p}_{\rm CM}\rangle \otimes |\Psi_{\rm re}\rangle$. Here, $|\Psi_{\rm re}\rangle$ is the energy eigenstate of the Hamiltonian for the relative coordinates
\begin{align}
        \hat{H}_{\rm re} |\Psi_{\rm re}\rangle = E_{\rm re} |\Psi_{\rm re}\rangle ,
\end{align}
and $|\Vec{p}_{\rm CM}\rangle$ is the plane wave eigenstate for the center of mass state. Note that $H_{\rm re}$ includes the potential of nucleus being at the origin by approximation. Therefore, the relative coordinates represent the positions of electrons. 

In the following, the energy eigenstate $|\Psi_{\rm re}\rangle$ is assumed to be given by the single Slater determinant of the single-electron wave function. Also, the single-electron wave functions are expressed by the spherical spinors (or spinor spherical harmonics) (see, e.g., Ref.~\cite{johnson2007atomic} for review).

\section{Interaction potential}
\label{sec:pot}

The Coulomb potential $\phi(\Vec{r})$ of atom is described by the Maxwell equation as 
\begin{align}
    \nabla^2 \phi(\Vec{r}) = - \left[Ze\delta(\Vec{r}-\Vec{r}_n) -e n(\Vec{r})\right] ,
    \label{eq:Poisson}
\end{align}
where $n(\Vec{r})$ is the number density of electrons in an atom and $e$ is the elementary charge. Assuming that the number density is written by the positions of each electron $n(\Vec{r}) = \sum_{i=1}^Z \delta(\Vec{r}-\Vec{r}_{e,i})$, the solution is 
\begin{align}
    \phi(\Vec{r}) = & \frac{Ze}{4\pi|\Vec{r}-\Vec{r}_n|} -\sum_{i=1}^Z \frac{e}{4\pi|\Vec{r}-\Vec{r}_{e,i}|} \\
    = & \int \frac{d^3\Vec{q}}{(2\pi)^3} \left( \frac{Ze}{q^2} e^{-i\Vec{q}\cdot(\Vec{r}-\Vec{r}_n)} - \sum_{i=1}^Z \frac{e}{q^2} e^{-i\Vec{q}\cdot(\Vec{r}-\Vec{r}_{e,i})} \right) .
\end{align}
The interaction with the nucleus can be ignored approximately when the excitation and ionization processes are considered. This is because the potential from the nucleus interactions depend on the coordinates as $\Vec{r} - \Vec{r}_n$ and $\Vec{r}_n$ is related to relative coordinates $\Vec{r}_i$ with the suppression factor of $\approx m_e/m_n$.

Then, the interaction Lagrangian for the ionization scattering of electron and atom is given by
\begin{align}
    \mathcal{L}_{\rm int} \approx e \bar{\psi}_e \gamma^\mu \psi_e A_\mu ,
\end{align}
where $\psi_e$ and $A_\mu$ are the electron and photon fields and $\gamma^\mu$ is the gamma matrix. The invariant scattering amplitude is given by the $t$-channel and $u$-channel processes as
\begin{align}
    \mathcal{M}_i \approx 16\pi \alpha m_e^2 \left( \frac{\delta_{s_{\rm in}^F s_{\rm in}^I}\delta_{s_i^F s_i^I}}{t_i} - \frac{\delta_{s_i^F s_{\rm in}^I}\delta_{s_{\rm in}^F s_i^I}}{u_i} \right)
\end{align}
under the non-relativistic approximation, where $t_i = (p_{{\rm in},I}-p_{{\rm in},F})^2$ and $u_i = (p_{{\rm in},I}-p_{e,i,F})^2$ are the  Mandelstam variables for $i$-th electron in the atom. The subscript $I$ and $F$ represent the label for initial and final states, $s$ represents the spin, and an incoming electron is labeled by in.

Then, the invariant amplitude is related to the interaction potential $V(\Vec{r}_{e,i}-\Vec{r}_{\rm in})$~\cite{sakurai1967advanced} by the Born approximation. The potential is written as 
\begin{align}
    V(\Vec{r}_{e,i}-\Vec{r}_{\rm in}) = V_t(\Vec{r}_{e,i}-\Vec{r}_{\rm in}) + V_u(\Vec{r}_{e,i}-\Vec{r}_{\rm in}) ,
\end{align}
where~\cite{Ibe:2017yqa}
\begin{align}
    V_t(\Vec{r}_{e,i}-\Vec{r}_{\rm in}) = & -\int \frac{d^3\Vec{q}}{(2\pi)^3} e^{i\Vec{q}\cdot(\Vec{r}_{e,i}-\Vec{r}_{\rm in})} \frac{\mathcal{M}_t(q^2)}{4m_e^2}  , \\
    V_u(\Vec{r}_{e,i}-\Vec{r}_{\rm in}) = & -\int \frac{d^3\Vec{q}_u}{(2\pi)^3} e^{i\Vec{q}_u\cdot(\Vec{r}_{e,i}-\Vec{r}_{\rm in})} \frac{\mathcal{M}_u(q_u^2)}{4m_e^2}
\end{align}
and the invariant scattering amplitude is given as
\begin{align}
    \mathcal{M}_t(q^2) = & -16\pi\alpha \frac{m_e^2}{q^2} , \\
    \mathcal{M}_u(q_u^2) = & 16\pi\alpha \frac{m_e^2}{q_u^2} .
\end{align}
Here, $s_i = (p_{{\rm in},I}+p_{e,i,I})^2$ is the Mandelstam variable, $\Vec{q}_u = \Vec{p}_{{\rm in},I}-\Vec{p}_{e,i,F}$ is the $u$-channel momentum transfer, $\Vec{q} = \Vec{p}_{{\rm in},I}-\Vec{p}_{{\rm in},F}$ is the momentum transfer of $t$-channel, and $\Vec{r}_{\rm in}$ is the coordinates of incoming electron. Remember that the $i$-th electron coordinate $\Vec{r}_{e,i}$ is expressed by the relative coordinate $\Vec{r}_i$ as $\Vec{r}_{e,i} = \Vec{r}_i + \Vec{r}_{\rm CM}$.

The above assumptions of a single Slater determinant and pure Coulomb potential neglect the electron correlation. These assumptions are not good in low energy scattering region and better in high energy region~\cite{peixoto1969elastic,naon1972importance,khatun2021theoretical,parvin2023theoretical}. This is because the incident electron passes through the atom in a short time. Ref.~\cite{peixoto1969elastic} shows that those approximations look good for $40$ keV of incident energy and the discrepancy is about 6-12 \% in small scattering angle region and better in large angle region for He and Ne atoms. In addition, this approach partially incorporates electron screening and correlation effects beyond the pure Coulomb potential and goes beyond a hydrogenic approximation.

\section{Electron-atom inelastic scattering cross section}
\label{sec:cross_section}

The initial state $|\Phi_I\rangle$ and final state $|\Phi_F\rangle$ for asymptotic states are given by $|\Phi_{I/F}\rangle = \sqrt{2m_A}\sqrt{2m_e}|\Psi_A\rangle_{I/F} |\vec{p}_{\rm in},s_{\rm in}\rangle_{I/F}$. Incoming electron states are taken as plane waves. Then, the $T$-matrix by a $t$-channel potential $V_t(\Vec{r}_i + \Vec{r}_{\rm CM}-\Vec{r}_{\rm in})$ becomes 
\begin{align}
    iT_{t,FI} = & -2\pi i \delta(E_F-E_I) \sum_{i=1}^Z \int d^3\Vec{r}_{\rm in} d^3\Vec{r}_{\rm CM} \left( \Pi_{j=1}^Z d^3\Vec{r}_{j} \right) \nonumber \\ 
    & \times \Phi_F^{\dagger} (\Vec{r}_{\rm in},\Vec{r}_{\rm CM},\{\Vec{r}_k\}) V_t(\Vec{r}_i + \Vec{r}_{\rm CM}-\Vec{r}_{\rm in}) \Phi_I (\Vec{r}_{\rm in},\Vec{r}_{\rm CM},\{\Vec{r}_l\}) \\
    = & i(2\pi)^4 \delta^4(p_F-p_I) \delta_{s_{\rm in}^Fs_{\rm in}^I} \frac{m_A}{m_e} \mathcal{M}_t(q^2) Z_{FI}(\Vec{q}) , 
    \label{eq:T}
\end{align}
where the transition matrix of single electron in atom $ Z_{FI}(\Vec{q})$ is defined as
\begin{align}
    Z_{FI}(\Vec{q}) \equiv & \sum_{i=1}^Z \int \left( \Pi_{j=1}^Z d^3\Vec{r}_{j} \right)  \Psi_{{\rm re},F}^\dagger (\{\Vec{r}_k\}) e^{i\Vec{q}\cdot \Vec{r}_i} \Psi_{{\rm re},I} (\{\Vec{r}_l\}) \\
    = & \int d^3\Vec{r} \phi_{E_{e,F}}^\dagger(\Vec{r}) e^{i\Vec{q}\cdot \Vec{r}} \phi_{E_{e,I}}(\Vec{r}) \\
    \approx & |\Vec{q}| z_{FI}(\theta_{qr}) .
    \label{eq:z}
\end{align}
Here, $\phi_{E_{e}}(\Vec{r})$ is the spherical spinor of single-electron energy $E_e$ and $E_e\approx \Vec{p}\,^2/(2 m_e)$ for the ionized state. The dipole approximation is used in Eq.~\eqref{eq:z} and the reduced transition matrix element is defined as 
\begin{align}
    z_{FI}(\theta_{qr}) \equiv i\int d^3\Vec{r} \phi_{E_{e,F}}^\dagger(\Vec{r})  |\Vec{r}|\cos\theta_{qr} \phi_{E_{e,I}}(\Vec{r}) ,
\end{align}
where $\theta_{qr}$ is the angle between $\Vec{q}$ and $\Vec{r}$, $\Vec{q} = -\Vec{p}_{{\rm in},F} + \Vec{p}_{{\rm in},I} = \Vec{p}_{{\rm CM},F} - \Vec{p}_{{\rm CM},I}$ is again the momentum transfer, $\delta^4(p_F-p_I) = \delta^4(p_{{\rm in},F} + p_{{\rm CM},F} - p_{{\rm in},I} - p_{{\rm CM},I})$ shows the four-momentum conservation, and $E_{I/F}$ is the total energy of the initial and final states. Here, this discussion does not use the sudden approximation of Migdal~\cite{migdal1941ionizacia,Landau:1991wop}, i.e., Galilei transformation of electrons at sudden time, because of the use of center of mass coordinates of atom. Therefore, this calculation is justified even for the case of the Coulomb force. This point is shown for the case of dark matter-nucleus scattering for hydrogen atom in Ref.~\cite{Kahn:2021ttr}. 

The $T$-matrix for $u$-channel potential becomes
\begin{align}
    iT_{u,FI} = & i(2\pi)^4 \delta(E_F-E_I) \frac{m_A}{m_e} \int \frac{d^3\Vec{q}_u}{(2\pi)^3} \delta^3(\Vec{q}_u+\Vec{p}_{{\rm CM},I}-\Vec{p}_{{\rm CM},u,F}) \mathcal{M}_u(q_u^2) \nonumber \\ 
    & \times \int d^3 \Vec{r}_{\rm in} \phi_{E_e,F}^\dagger(\Vec{r}_{\rm in}) e^{-i\Vec{q}_u\cdot\Vec{r}_{\rm in}} \langle \Vec{r}_{\rm in}|\Vec{p}_{{\rm in},I}, s_{{\rm in},I} \rangle \nonumber \\
    & \times \int d^3 \Vec{r} \langle \Vec{p}_{{\rm in},F}, s_{{\rm in},F}| \Vec{r} \rangle e^{i\Vec{q}_u\cdot\Vec{r}} \phi_{E_e,I}(\vec{r}) .
\end{align}
Here, $\Vec{p}_{{\rm CM},u,F} = \Vec{p}_{{\rm CM},F} - \Vec{p}_{i,F} + \Vec{p}_{{\rm in},F}$ because of the $u$-channel process. The integrations by $\Vec{r}_{\rm in}$ and $\Vec{r}$ are actually the Fourier transformations of the projected single-electron wave functions. Assuming that the plane wave $\langle \Vec{r}\,| \Vec{p}, s \rangle$ is approximately equal to the Coulomb wave $\phi (\Vec{r})$ (see Ref.~\cite{gozem2015photoelectron} for the discussion of the differences), the above equation becomes 
\begin{align}
    iT_{u,FI} \approx & i(2\pi)^4 \delta(E_F-E_I) \delta^3(\Vec{p}_{{\rm in},I}+\Vec{p}_{{\rm CM},I}-\Vec{p}_{{\rm in},F}-\Vec{p}_{{\rm CM},F}) \delta_{s_i^Fs_{\rm in}^I} \frac{m_A}{m_e}  \mathcal{M}_u(q_u^2) Z_{{\rm in}I}(\Vec{q}_u) .
\end{align}
Here, the transition matrix is given as
\begin{align}
    Z_{{\rm in}I}(\Vec{q}_u) = \int d^3 \Vec{r} \phi_{{\rm in},F}^\dagger(\Vec{r}) e^{i\Vec{q}_u\cdot\Vec{r}} \phi_{E_e,I}(\Vec{r}) ,
\end{align}
where $\phi_{{\rm in},F}(\Vec{r})$ is the Coulomb wave function with energy for incoming electron and $\Vec{q}_u = \Vec{p}_{{\rm in},I} - \Vec{p}_{i,F} = \Vec{p}_{{\rm CM},F} - \Vec{p}_{{\rm CM},I} - \Vec{p}_{i,F} + \Vec{p}_{{\rm in},F}$ is again the momentum transfer for $u$-channel process. This approximation will be not good for excitation processes because the excitation states are actually bound states. 

Finally, the $T$-matrix is obtained as
\begin{align}
    iT_{FI} = & iT_{t,FI} + iT_{u,FI} \\
    = & i(2\pi)^4 \delta^4(p_F-p_I) \frac{m_A}{m_e} \left( \delta_{s_{\rm in}^Fs_{\rm in}^I} \mathcal{M}_t(q^2) Z_{FI}(\Vec{q}) + \delta_{s_i^Fs_{\rm in}^I}   \mathcal{M}_u(q_u^2) Z_{{\rm in}I}(\Vec{q}_u) \right) .
\end{align}

The four-momentum in the laboratory frame is expressed by $p_{{\rm in},I} = (m_e + E_{{\rm kin},I}, \Vec{p}_{{\rm in},I})$, $p_{{\rm in},F} = (m_e + E_{{\rm kin},F},\Vec{p}_{{\rm in},F})$, $p_{{\rm CM},I} = (m_A, 0)$, and $p_{{\rm CM},F} = (E_{A,F}, \Vec{p}_{A,F})$, where $E_{\rm kin} \approx \Vec{p}_{\rm in}\,^2/(2m_e)$ is the kinetic energy of incoming electron in the laboratory frame. The momentum of final state atom becomes $\Vec{p}_{A,F} \approx \Vec{p}_{n,F} + \Vec{p}_{i,F}$. Here, $\Vec{p}_{i,F}=0$ for excitation. Then, $m_A = m_n + Z m_e + E_{{\rm re},I}$ and $E_{A,F} \approx m_n + Z m_e + E_{{\rm re},F} + \Vec{p}_{n,F}\,^2/(2 m_n)$ are followed. Finally, the minimum and maximum momentum transfer, $q^2_{\rm min} \approx 2m_e \Delta E$ and $q^2_{\rm max} \approx 2m_e (E_{{\rm kin},I}-\Delta E)$, are obtained, where $\Delta E = E_{e,F}-E_{e,I}>0$ is the transition energy of electron in the atom. The $u$-channel momentum transfer becomes $\Vec{q}_u\,^2 \approx 2m_e E_{{\rm kin},I}-\Vec{q}\,^2$.

The total inelastic cross section of single-electron scattering is obtained as 
\begin{align}
    \sigma(\Vec{p}_{{\rm in},I}\,^2) = & \frac{1}{2!} \frac{1}{2}\sum_{F,s_{\rm in}^{F/I}} \int \frac{d^3\Vec{p}_{{\rm in},F}}{(2\pi)^3 2 p_{{\rm in},F}^0} \frac{d^3\Vec{p}_{{\rm CM},F}}{(2\pi)^3 2 p_{{\rm CM},F}^0} \frac{1}{4\sqrt{(p_{{\rm CM},I}\cdot p_{{\rm in},I})^2-m_A^2 m_e^2}} \nonumber \\ 
    & \times \left(\frac{m_A}{m_e}\right)^2 (2\pi)^4 \delta^4(p_{{\rm in},F} + p_{{\rm CM},F} - p_{{\rm in},I} - p_{{\rm CM},I}) \nonumber \\
    & \times \left| \delta_{s_{\rm in}^Fs_{\rm in}^I} \mathcal{M}_t(q^2) Z_{FI}(\Vec{q}) + \delta_{s_i^Fs_{\rm in}^I}   \mathcal{M}_u(q_u^2) Z_{{\rm in}I}(\Vec{q}_u) \right|^2 .
\end{align}
In the cross section, the possible final states $F$ and spins are summed or averaged. The electron spins of final states $s_i^F$ are included in the label $F$. The sum of the initial spin for atomic electrons are included in $Z_{FI}$~\cite{Ibe:2017yqa}. Note that the sum includes the integration of ionized electron energy. This equation is reduced as
\begin{align}
    \sigma(\Vec{p}_{{\rm in},I}\,^2) = & \frac{4\pi \alpha^2 m_e^2}{\Vec{p}_{{\rm in},I}\,^2} \sum_{F} \left[ \left|z_{FI}\right|^2 \log \frac{q^2_{\rm max}}{q^2_{\rm min}} + \left|z_{{\rm in}I}\right|^2 \log \frac{2m_e E_{{\rm kin},I}-q_{\rm min}^2}{2m_e E_{{\rm kin},I}-q_{\rm max}^2} \right. \nonumber \\
    & \left.-2 {\rm Re}\left[ z_{FI} z_{{\rm in}I}^* \right] \left( \sin^{-1} \sqrt{\frac{q_{\rm max}^2}{2m_e E_{{\rm kin},I}}} - \sin^{-1} \sqrt{\frac{q_{\rm min}^2}{2m_e E_{{\rm kin},I}}} \right) \right] ,
    \label{eq:sigma_p}
\end{align}
or, expressed by the kinetic energy of incoming electron $E_{{\rm kin},I}$ as  
\begin{align}
    \sigma(E_{{\rm kin},I}) = & \frac{2\pi \alpha^2 m_e}{E_{{\rm kin},I}} \sum_{F} \left[ \left(\left|z_{FI}\right|^2 + \left|z_{{\rm in}I}\right|^2 \right) \log \frac{E_{{\rm kin},I}-\Delta E}{\Delta E} \right. \nonumber \\
    & \left.-2 {\rm Re}\left[ z_{FI} z_{{\rm in}I}^* \right] \left( \sin^{-1} \sqrt{1-\frac{\Delta E}{E_{{\rm kin},I}}} - \sin^{-1} \sqrt{\frac{\Delta E}{E_{{\rm kin},I}}} \right) \right] .
    \label{eq:sigma_E}
\end{align}
To obtain Eq.~\eqref{eq:sigma_p}, the average/sum of magnetic quantum number for the initial/final states are taken, which makes the factor $\left|z_{FI}(\theta_{qr})\right|^2$ isotropic for the angular direction $\left|z_{FI}(\theta_{qr})\right|^2\to \left|z_{FI}\right|^2$~\cite{Ibe:2017yqa}. Also, the fact that the transition matrix element does not change the spins is used. The final states of $z_{FI} z_{{\rm in}I}^*$ are related by the energy conservation as $E_{{\rm in},F} \approx E_{{\rm kin},I} - \Delta E$. The log term shows the kinematical constraint from the phase space of $E_{e,F}$ as $E_{{\rm kin},I}-2\Delta E > 0$. The $\left|z_{FI}\right|^2$ part in the first term is actually the scattering cross section by the Coulomb potential without the identification factor of the interacting electrons $1/2!$.

The averaged value of $|z_{FI}|^2$ is actually the transition probabilities for the Migdal effect given in Ref.~\cite{Ibe:2017yqa} which uses the \texttt{Flexible Atomic Code (FAC, cFAC)}~\cite{doi:10.1139/p07-197} for the calculation. \texttt{cFAC} calculates the probabilities with the Dirac-Hartree-Fock method and a universal central potential.

\begin{figure}[t]
    \centering
    \includegraphics[width=0.6\textwidth]{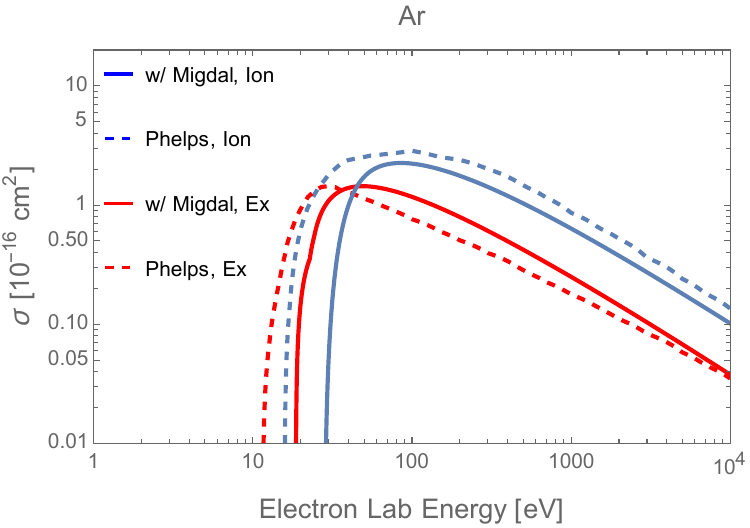} \\[4ex]
    \includegraphics[width=0.6\textwidth]{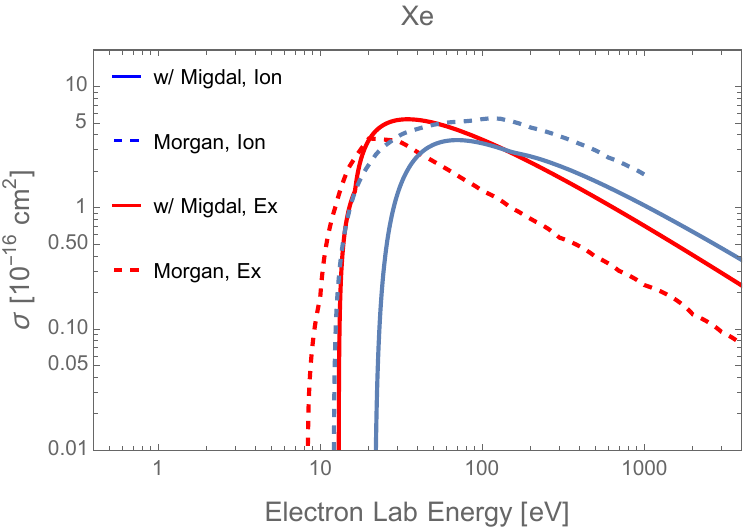}
    \caption{
    The total excitation and ionization cross section of electron-atom scattering $\sigma(E_{{\rm kin},I})$ as the function of kinetic energy for incoming electron in laboratory frame $E_{{\rm kin},I}$. The calculation results using the transition probabilities of Ref.~\cite{Ibe:2017yqa} are shown by the solid lines and the comparison values are shown by the dashed lines. The red lines show the single excitation case and blue lines show the single ionization case. 
    (top:) For Ar case. The database value for comparison is taken from Phelps~\cite{Phelps:2024}.
    (bottom:) For Xe case. The database value for comparison is taken from Morgan~\cite{Morgan:2024}. The ionization data is limited up to 1 keV.
 }
 \label{fig:cross_section}
\end{figure}  

The calculation result of $\sigma(E_{{\rm kin},I})$ is shown in Fig.~\ref{fig:cross_section}. For Ar, the difference of ionization cross section is around $30\ \%$ from the database value in 5 keV $\lesssim E_{{\rm kin},I} < 10$ keV. That of excitation has local maximum of around $60\ \%$ at 100 eV and gradually decreasing by increasing $E_{{\rm kin},I}$. For Xe, the difference of ionization cross section is around 80 \% at 1 keV and that of excitation is roughly factor 3 at 1 keV. This gap dependency is consistent with that the Born approximation is a good approximation in the keV energy region~\cite{bonham1974high,garcia1999electron}. The interference term is effective at low energy and negligible at high energy. Note that the probability is almost dominated by the excitation/ionization from the outermost shell. For Ar, $3p$ states dominate the contribution and $2p$ and $3s$ states give $\sim 1\ \%$ contribution at $E_{{\rm kin},I} = 10$ keV. For Xe, $5p$ states dominate up to $\sim$ 100 eV, but, $4d$ states have $\sim 40\ \%$ contribution at $E_{{\rm kin},I} = 1$ keV.

\section{Comparison with other results}

The inelastic cross section is also calculated by using the Bethe theory~\cite{inokuti1971inelastic,inokuti1975total,inokuti1981oscillator,salvat2022inelastic}. The obtained formula of inelastic cross section for Ar is~\cite{inokuti1975total}
\begin{align}
    \sigma_{\rm Inokuti} (E_{{\rm kin},I}) = 4\pi a_0^2 \frac{\rm Ryd}{E_{{\rm kin},I}} \left[ M_{\rm Inokuti}^2 \log\left(4\frac{E_{{\rm kin},I}}{\rm Ryd}\right)+M_{\rm Inokuti}^2\log c_{\rm Inokuti} + \gamma_{\rm Inokuti}\frac{\rm Ryd}{E_{{\rm kin},I}} \right] ,
\end{align}
where $a_0 = 1/(m_e \alpha)$ is the Bohr radius, Ryd = 13.605 eV is the Rydberg energy, $\gamma_{\rm Inokuti} = Z(-7/4 +\log (B/E_{{\rm kin},I}))$ for electron, and $B$ is the average binding energy which is taken as 13 eV because the 3p states dominate the signal. $M_{\rm Inokuti}^2 = 5.4755$ and $M_{\rm Inokuti}^2\log c_{\rm Inokuti} = -1.063$ are taken from the Hartree-Fock calculation (see Ref.~\cite{inokuti1975total}). Ref.~\cite{inokuti1975total,inokuti1981oscillator} calculated the ionization cross section from hydrogen ($Z = 1$) to strontium ($Z = 38$).

For Xe, the result of Ref.~\cite{salvat2022inelastic} is used. It shows
\begin{align}
    \sigma_{\rm Salvat} (\beta^2) = \frac{2\pi\alpha^2}{m_e \beta^2} \left[ M_{\rm Salvat}^2\left( \log \frac{\beta^2}{1-\beta^2} -\beta^2 \right) + C_{\rm Salvat} \right] ,
\end{align}
where $\beta$ is the velocity of incoming electron and the values of $M_{\rm Salvat}^2$ and $C_{\rm Salvat}$ are taken from the figure in Ref.~\cite{salvat2022inelastic}. This calculation used the shell corrections, self-consistent Dirac-Hartree-Fock-Slater potential, independent-electron approximation, and relativistic plane-wave Born approximation. They reported less than 1 \% precision for $E \gtrsim 100$ MeV and $\sim 8$ \% precision for $E \sim 5$ MeV for high $Z$ atoms and good agreement with the results of Ref.~\cite{inokuti1975total,inokuti1981oscillator}.

\begin{figure}[t]
    \centering
    \includegraphics[width=0.6\textwidth]{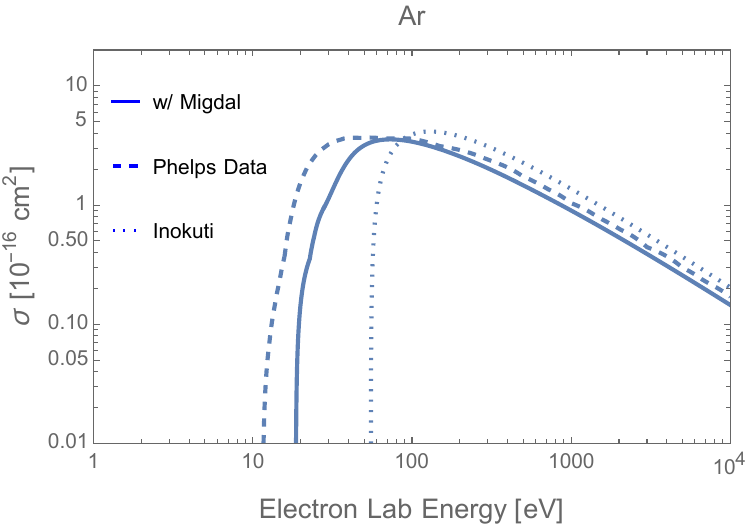} \\[4ex]
    \includegraphics[width=0.6\textwidth]{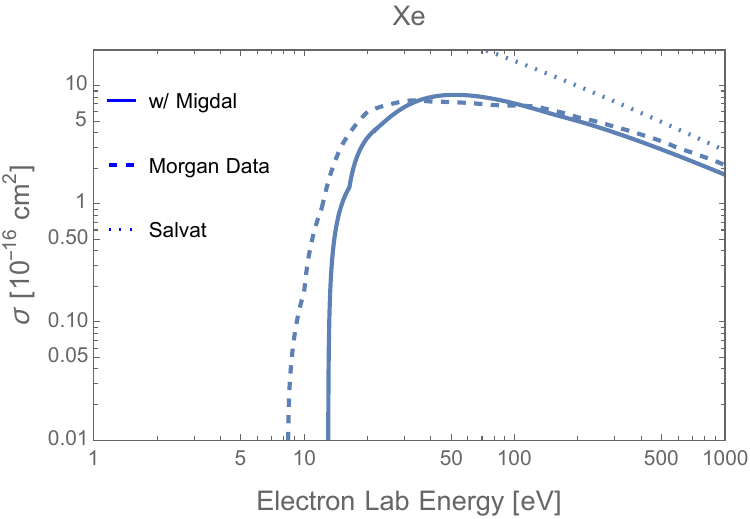}
    \caption{
    The comparison of total inelastic cross section $\sigma(E_{{\rm kin},I})$. The calculation results using Ref.~\cite{Ibe:2017yqa} are shown by the solid lines and the comparison values are shown by the dashed and dotted lines. 
    (top:) For Ar case. The dotted lines are from the theoretical calculation in Ref.~\cite{inokuti1975total} with the Hartree-Fock calculation. The database value is taken from Phelps~\cite{Phelps:2024}.
    (bottom:) For Xe case. The dotted lines are from the theoretical calculation in Ref.~\cite{salvat2022inelastic} and not convergent at low energy region because of its formula. The database value is taken from Morgan~\cite{Morgan:2024}. 
 }
 \label{fig:rel}
\end{figure}  

The comparison result is shown in Fig.~\ref{fig:rel}. For both comparisons, this result shows smaller value than the database value and Ref.~\cite{inokuti1975total,salvat2022inelastic}. Also, the result from Ref.~\cite{salvat2022inelastic} gets consistent parameters with that from Ref.~\cite{inokuti1975total}. The line of Ref.~\cite{salvat2022inelastic} aims over MeV region with heavy incoming particles and does not converge at low energy region because of its shape of formula. Their result will be applied here because electrons are not fast in keV.

The binding energies and wave functions of electrons are calculated with electron correlations, e.g., Ref.~\cite{Baxter:2019pnz} shows $\sim 5$ \% precision for binding energies which is better than Ref.~\cite{Ibe:2017yqa}. Also, Ref.~\cite{Cox:2022ekg} shows $\sim 20$ \% differences of ionization probabilities for xenon 5s and argon 3p states with a 20 keV ionized electron. These uncertainties are significant for outer shell states and this comparison method actually relies on the precision of outermost shell states. Note that the uncertainty becomes much smaller for inner shell electrons. Density functional theory may provide better precision of binding energies, but, it does not provide the accurate wave functions.

In addition, this scattering formulation approximates the electron correlations, but this uncertainty is difficult to estimate. 
The use of $u$-channel potential will be valid for $E_{{\rm kin},I} \gtrsim 1$ keV electron because faster electrons feel less effect of atomic electron clouds than slower ones.

\section{Discussion and conclusion}

The excitation/ionization probability from dark matter-nucleus scattering is compared to the electron-atom inelastic cross section. The differences are $\sim 30\ \%$ for Ar at 10 keV and $\sim 80\ \%$ for Xe compared with the database value shown in Fig.~\ref{fig:cross_section}. For Xe, the database value is limited and the differences may be improved for higher energy because the plane wave and the Coulomb wave approximation used for $u$-channel processes becomes better for high energy. This approximation is not good for excitation cross sections because the excited states are approximated to the plane wave states. Therefore, the expected differences for excitation are not reliable. The differences in the low energy threshold come from the kinematical constraint, $E_{{\rm kin},I}>2\Delta E$. In addition, the calculated bound energies are reported to have $\mathcal{O}(10)\ \%$ differences from the experimental data~\cite{Ibe:2017yqa}. 
In this calculation, the change of the electron mass in $m_A$ by scattering is neglected. This change will cause a difference of $m_e/m_n \sim 10^{-4}$.

The electron-nucleus interaction will also cause the excitation/ionization of electrons in small probability. This effect corresponds to the Migdal effect for the Coulomb force, which is neglected in this study because the factor $m_e/m_n$ suppression is expected in amplitude. Also, there are excitation/ionization processes of $j$-th electron from the interaction with $i$-th electron ($j\ne i$).

An isolated atom is assumed in this calculation, but large volume detectors use the liquid Xe. For electron, the effect of liquid structure becomes effective for $E_{{\rm kin},I} \lesssim 10$ eV and the cross section corresponds to that of gas probes by increasing $E_{{\rm kin},I}$~\cite{atrazhev1981hot,atrazhev1985heating}. Then, for high energy electrons $E_{{\rm kin},I} \gg 10$ eV, the cross section will not so differ from the isolated atom calculation. But, the structure of outermost shell may differ between liquid and isolated probes.

This treatment of the Migdal effect in Ref.~\cite{Ibe:2017yqa} does not include explicit couplings between different electronic channels due to non-adiabatic transitions. The dimensionless order parameter is assumed from the time-dependent perturbation theory as $\kappa_{ab} = |\vec{d}_{ab}\cdot \vec{v}_n|/E_{ab}$. Here, $\vec{d}_{ab}$ is the transition matrix of states $a$ and $b$, $\vec{v}_n$ is the velocity of nuclear recoil, and $E_{ab}$ is the energy difference. Assuming $|\vec{d}_{ab}\cdot \vec{v}_n| \sim a_0^{-1} |\vec{v}_n|$, the parameter becomes $\kappa_{ab} \sim (7 \ {\rm keV}/E_{ab})\sqrt{E_R/m_n}$. Then, $\kappa_{ab} \ll 1$ will be satisfied at least for $E_R \lesssim 1$ keV or $E_{ab} \gtrsim 100$ eV. A fully consistent treatment of the Migdal effect including multi-channel non-adiabatic couplings and time-dependent correlation effects would require a more sophisticated quantum-chemistry or time-dependent multi-electron approach, which will be a future improvement.

\section*{Acknowledgements}

This work is supported by JSPS Grant-in-Aid for
Scientific Research
KAKENHI Grant No.~JP23KJ2173.

\appendix

\bibliographystyle{jhep}

\bibliography{main}

\providecommand{\href}[2]{#2}\begingroup\raggedright\begin{thebibliography}{100}

\bibitem{Murayama:2007ek}
H.~Murayama, \emph{{Physics Beyond the Standard Model and Dark Matter}},  in
  \emph{{Les Houches Summer School - Session 86: Particle Physics and
  Cosmology: The Fabric of Spacetime}}, 4, 2007,
  \href{https://arxiv.org/abs/0704.2276}{{\ttfamily 0704.2276}}.

\bibitem{Goodman:1984dc}
M.~W. Goodman and E.~Witten, \emph{{Detectability of Certain Dark Matter
  Candidates}}, \href{https://doi.org/10.1103/PhysRevD.31.3059}{\emph{Phys.
  Rev. D} {\bfseries 31} (1985) 3059}.

\bibitem{Lee:1977ua}
B.~W. Lee and S.~Weinberg, \emph{{Cosmological Lower Bound on Heavy Neutrino
  Masses}}, \href{https://doi.org/10.1103/PhysRevLett.39.165}{\emph{Phys. Rev.
  Lett.} {\bfseries 39} (1977) 165}.

\bibitem{Lewin:1995rx}
J.~D. Lewin and P.~F. Smith, \emph{{Review of mathematics, numerical factors,
  and corrections for dark matter experiments based on elastic nuclear
  recoil}},
  \href{https://doi.org/10.1016/S0927-6505(96)00047-3}{\emph{Astropart. Phys.}
  {\bfseries 6} (1996) 87}.

\bibitem{migdal1939ionizatsiya}
A.~Migdal, \emph{Ionizatsiya atomov pri yadernykh reaktsiyakh}, {\emph{Sov.
  Phys. JETP} {\bfseries 9} (1939) 1163}.

\bibitem{migdal1941ionizacia}
A.~Migdal, \emph{Ionizacia atomov v $\alpha$-i $\beta$-raspade}, {\emph{ZhETF}
  {\bfseries 11} (1941) 207}.

\bibitem{Landau:1991wop}
L.~D. Landau and E.~M. Lifshits, \emph{{Quantum Mechanics}: {Non-Relativistic
  Theory}}, vol.~v.3 of \emph{Course of Theoretical Physics}.
  Butterworth-Heinemann, Oxford, 1991.

\bibitem{Rapaport:1975zza}
M.~S. Rapaport, F.~Asaro and I.~Perlman, \emph{{K-shell electron shake-off
  accompanying alpha decay}},
  \href{https://doi.org/10.1103/PhysRevC.11.1740}{\emph{Phys. Rev. C}
  {\bfseries 11} (1975) 1740}.

\bibitem{Rapaport:1975zz}
M.~S. Rapaport, F.~Asaro and I.~Perlman, \emph{{L- and M-shell electron
  shake-off accompanying alpha decay}},
  \href{https://doi.org/10.1103/PhysRevC.11.1746}{\emph{Phys. Rev. C}
  {\bfseries 11} (1975) 1746}.

\bibitem{fischbeck1975spectroscopy}
H.~Fischbeck and M.~Freedman, \emph{Spectroscopy of $\alpha$ and k-and
  l-electron continua and l-electron pickup in po 210 $\alpha$ decay},
  {\emph{Physical Review Letters} {\bfseries 34} (1975) 173}.

\bibitem{fischbeck1977angular}
H.~Fischbeck and M.~Freedman, \emph{Angular correlation between ejected l
  electrons and $\alpha$ particles in po 210 decay}, {\emph{Physical Review A}
  {\bfseries 15} (1977) 162}.

\bibitem{boehm1954internal}
F.~Boehm and C.~Wu, \emph{Internal bremsstrahlung and ionization accompanying
  beta decay}, {\emph{Physical Review} {\bfseries 93} (1954) 518}.

\bibitem{berlovich1965investigation}
{\'E}.~Berlovich, L.~Kutsentov and V.~Fle^^c7^^90sher, \emph{Investigation of
  the" jolting" of electron shells of oriented molecules containing p 32},
  {\emph{Soviet Journal of Experimental and Theoretical Physics} {\bfseries 21}
  (1965) 675}.

\bibitem{couratin2012first}
C.~Couratin, P.~Velten, X.~Fl{\'e}chard, E.~Li{\'e}nard, G.~Ban, A.~Cassimi
  et~al., \emph{First measurement of pure electron shakeoff in the $\beta$
  decay of trapped he+ 6 ions}, {\emph{Physical Review Letters} {\bfseries 108}
  (2012) 243201}.

\bibitem{Lienard:2015joa}
E.~Li\'enard et~al., \emph{{Precision measurements with LPCTrap at GANIL}},
  \href{https://doi.org/10.1007/978-3-319-61588-2_1}{\emph{Hyperfine Interact.}
  {\bfseries 236} (2015) 1} [\href{https://arxiv.org/abs/1507.05838}{{\ttfamily
  1507.05838}}].

\bibitem{Fabian:2018pin}
X.~Fabian et~al., \emph{{Electron shakeoff following the ${\beta}^{+}$ decay of
  $^{19}\mathrm{Ne}^{+}$ and $^{35}\mathrm{Ar}^{+}$ trapped ions}},
  \href{https://doi.org/10.1103/PhysRevA.97.023402}{\emph{Phys. Rev. A}
  {\bfseries 97} (2018) 023402}
  [\href{https://arxiv.org/abs/1802.01298}{{\ttfamily 1802.01298}}].

\bibitem{Nakamura:2020kex}
K.~D. Nakamura, K.~Miuchi, S.~Kazama, Y.~Shoji, M.~Ibe and W.~Nakano,
  \emph{{Detection capability of the Migdal effect for argon and xenon nuclei
  with position-sensitive gaseous detectors}},
  \href{https://doi.org/10.1093/ptep/ptaa162}{\emph{PTEP} {\bfseries 2021}
  (2021) 013C01} [\href{https://arxiv.org/abs/2009.05939}{{\ttfamily
  2009.05939}}].

\bibitem{Araujo:2022wjh}
H.~M. Ara\'ujo et~al., \emph{{The MIGDAL experiment: Measuring a rare atomic
  process to aid the search for dark matter}},
  \href{https://doi.org/10.1016/j.astropartphys.2023.102853}{\emph{Astropart.
  Phys.} {\bfseries 151} (2023) 102853}
  [\href{https://arxiv.org/abs/2207.08284}{{\ttfamily 2207.08284}}].

\bibitem{baur1983ionisation}
G.~Baur, F.~Rosel and D.~Trautmann, \emph{Ionisation induced by neutrons},
  {\emph{Journal of Physics B: Atomic and Molecular Physics} {\bfseries 16}
  (1983) L419}.

\bibitem{vegh1983multiple}
L.~Vegh, \emph{Multiple ionisation effects due to recoil in atomic collisions},
  {\emph{Journal of Physics B: Atomic and Molecular Physics} {\bfseries 16}
  (1983) 4175}.

\bibitem{wauters1997recoil}
L.~Wauters, N.~Vaeck, M.~Godefroid, H.~W. Van Der~Hart and M.~Demeur,
  \emph{Recoil-induced electronic excitation and ionization in one-and
  two-electron ions}, {\emph{Journal of Physics B: Atomic, Molecular and
  Optical Physics} {\bfseries 30} (1997) 4569}.

\bibitem{berakdar2001electronic}
J.~Berakdar, \emph{Electronic correlation studied by neutron scattering},
  {\emph{Journal of Physics B: Atomic, Molecular and Optical Physics}
  {\bfseries 35} (2001) L31}.

\bibitem{Vergados:2005dpd}
J.~D. Vergados and H.~Ejiri, \emph{{The role of ionization electrons in direct
  neutralino detection}},
  \href{https://doi.org/10.1016/j.physletb.2004.11.085}{\emph{Phys. Lett. B}
  {\bfseries 606} (2005) 313}
  [\href{https://arxiv.org/abs/hep-ph/0401151}{{\ttfamily hep-ph/0401151}}].

\bibitem{Moustakidis:2005gx}
C.~C. Moustakidis, J.~D. Vergados and H.~Ejiri, \emph{{Direct dark matter
  detection by observing electrons produced in neutralino-nucleus collisions}},
  \href{https://doi.org/10.1016/j.nuclphysb.2005.08.033}{\emph{Nucl. Phys. B}
  {\bfseries 727} (2005) 406}
  [\href{https://arxiv.org/abs/hep-ph/0507123}{{\ttfamily hep-ph/0507123}}].

\bibitem{Ejiri:2005aj}
H.~Ejiri, C.~C. Moustakidis and J.~D. Vergados, \emph{{Dark matter search by
  exclusive studies of X-rays following WIMPs nuclear interactions}},
  \href{https://doi.org/10.1016/j.physletb.2006.03.037}{\emph{Phys. Lett. B}
  {\bfseries 639} (2006) 218}
  [\href{https://arxiv.org/abs/hep-ph/0510042}{{\ttfamily hep-ph/0510042}}].

\bibitem{talman2006excitations}
J.~D. Talman and A.~M. Frolov, \emph{Excitations of light atoms during nuclear
  reactions with fast neutrons}, {\emph{Physical Review A―Atomic, Molecular,
  and Optical Physics} {\bfseries 73} (2006) 032722}.

\bibitem{Bernabei:2007jz}
R.~Bernabei et~al., \emph{{On electromagnetic contributions in WIMP quests}},
  \href{https://doi.org/10.1142/S0217751X07037093}{\emph{Int. J. Mod. Phys. A}
  {\bfseries 22} (2007) 3155}
  [\href{https://arxiv.org/abs/0706.1421}{{\ttfamily 0706.1421}}].

\bibitem{liertzer2012multielectron}
M.~Liertzer, J.~Feist, S.~Nagele and J.~Burgd{\"o}rfer, \emph{Multielectron
  transitions induced by neutron impact on helium}, {\emph{Physical Review
  Letters} {\bfseries 109} (2012) 013201}.

\bibitem{Vergados:2013raa}
J.~D. Vergados, H.~Ejiri and K.~G. Savvidy, \emph{{Theoretical direct WIMP
  detection rates for inelastic scattering to excited states}},
  \href{https://doi.org/10.1016/j.nuclphysb.2013.09.010}{\emph{Nucl. Phys. B}
  {\bfseries 877} (2013) 36} [\href{https://arxiv.org/abs/1307.4713}{{\ttfamily
  1307.4713}}].

\bibitem{pindzola2014neutron}
M.~Pindzola, T.~Lee, S.~A. Abdel-Naby, F.~Robicheaux, J.~Colgan and M.~F.
  Ciappina, \emph{Neutron-impact ionization of he}, {\emph{Journal of Physics
  B: Atomic, Molecular and Optical Physics} {\bfseries 47} (2014) 195202}.

\bibitem{Sharma:2017fmo}
P.~Sharma, \emph{{Role of nuclear charge change and nuclear recoil on shaking
  processes and their possible implication on physical processes}},
  \href{https://doi.org/10.1016/j.nuclphysa.2017.08.004}{\emph{Nucl. Phys. A}
  {\bfseries 968} (2017) 326}.

\bibitem{Ibe:2017yqa}
M.~Ibe, W.~Nakano, Y.~Shoji and K.~Suzuki, \emph{{Migdal Effect in Dark Matter
  Direct Detection Experiments}},
  \href{https://doi.org/10.1007/JHEP03(2018)194}{\emph{JHEP} {\bfseries 03}
  (2018) 194} [\href{https://arxiv.org/abs/1707.07258}{{\ttfamily
  1707.07258}}].

\bibitem{Dolan:2017xbu}
M.~J. Dolan, F.~Kahlhoefer and C.~McCabe, \emph{{Directly detecting sub-GeV
  dark matter with electrons from nuclear scattering}},
  \href{https://doi.org/10.1103/PhysRevLett.121.101801}{\emph{Phys. Rev. Lett.}
  {\bfseries 121} (2018) 101801}
  [\href{https://arxiv.org/abs/1711.09906}{{\ttfamily 1711.09906}}].

\bibitem{LUX:2018akb}
{\scshape LUX} collaboration, \emph{{Results of a Search for Sub-GeV Dark
  Matter Using 2013 LUX Data}},
  \href{https://doi.org/10.1103/PhysRevLett.122.131301}{\emph{Phys. Rev. Lett.}
  {\bfseries 122} (2019) 131301}
  [\href{https://arxiv.org/abs/1811.11241}{{\ttfamily 1811.11241}}].

\bibitem{Bell:2019egg}
N.~F. Bell, J.~B. Dent, J.~L. Newstead, S.~Sabharwal and T.~J. Weiler,
  \emph{{Migdal effect and photon bremsstrahlung in effective field theories of
  dark matter direct detection and coherent elastic neutrino-nucleus
  scattering}}, \href{https://doi.org/10.1103/PhysRevD.101.015012}{\emph{Phys.
  Rev. D} {\bfseries 101} (2020) 015012}
  [\href{https://arxiv.org/abs/1905.00046}{{\ttfamily 1905.00046}}].

\bibitem{GrillidiCortona:2020owp}
G.~Grilli~di Cortona, A.~Messina and S.~Piacentini, \emph{{Migdal effect and
  photon Bremsstrahlung: improving the sensitivity to light dark matter of
  liquid argon experiments}},
  \href{https://doi.org/10.1007/JHEP11(2020)034}{\emph{JHEP} {\bfseries 11}
  (2020) 034} [\href{https://arxiv.org/abs/2006.02453}{{\ttfamily
  2006.02453}}].

\bibitem{vonKrosigk:2020udi}
B.~von Krosigk, M.~J. Wilson, C.~Stanford, B.~Cabrera, R.~Calkins, D.~Jardin
  et~al., \emph{{Effect on dark matter exclusion limits from new silicon
  photoelectric absorption measurements}},
  \href{https://doi.org/10.1103/PhysRevD.104.063002}{\emph{Phys. Rev. D}
  {\bfseries 104} (2021) 063002}
  [\href{https://arxiv.org/abs/2010.15874}{{\ttfamily 2010.15874}}].

\bibitem{XMASS:2022tkr}
{\scshape XMASS} collaboration, \emph{{Direct dark matter searches with the
  full data set of XMASS-I}},
  \href{https://doi.org/10.1103/PhysRevD.108.083022}{\emph{Phys. Rev. D}
  {\bfseries 108} (2023) 083022}
  [\href{https://arxiv.org/abs/2211.06204}{{\ttfamily 2211.06204}}].

\bibitem{Coskuner:2018are}
A.~Coskuner, D.~M. Grabowska, S.~Knapen and K.~M. Zurek, \emph{{Direct
  Detection of Bound States of Asymmetric Dark Matter}},
  \href{https://doi.org/10.1103/PhysRevD.100.035025}{\emph{Phys. Rev. D}
  {\bfseries 100} (2019) 035025}
  [\href{https://arxiv.org/abs/1812.07573}{{\ttfamily 1812.07573}}].

\bibitem{EDELWEISS:2019vjv}
{\scshape EDELWEISS} collaboration, \emph{{Searching for low-mass dark matter
  particles with a massive Ge bolometer operated above-ground}},
  \href{https://doi.org/10.1103/PhysRevD.99.082003}{\emph{Phys. Rev. D}
  {\bfseries 99} (2019) 082003}
  [\href{https://arxiv.org/abs/1901.03588}{{\ttfamily 1901.03588}}].

\bibitem{Tvrznikova:2019tgx}
L.~Tvrznikova, \emph{{Sub-GeV Dark Matter Searches and Electric Field Studies
  for the LUX and LZ Experiments}}, Ph.D. thesis, Yale U., 2019.
\newblock \href{https://arxiv.org/abs/1904.08979}{{\ttfamily 1904.08979}}.

\bibitem{CDEX:2019hzn}
{\scshape CDEX} collaboration, \emph{{Constraints on Spin-Independent Nucleus
  Scattering with sub-GeV Weakly Interacting Massive Particle Dark Matter from
  the CDEX-1B Experiment at the China Jinping Underground Laboratory}},
  \href{https://doi.org/10.1103/PhysRevLett.123.161301}{\emph{Phys. Rev. Lett.}
  {\bfseries 123} (2019) 161301}
  [\href{https://arxiv.org/abs/1905.00354}{{\ttfamily 1905.00354}}].

\bibitem{XENON:2019zpr}
{\scshape XENON} collaboration, \emph{{Search for Light Dark Matter
  Interactions Enhanced by the Migdal Effect or Bremsstrahlung in XENON1T}},
  \href{https://doi.org/10.1103/PhysRevLett.123.241803}{\emph{Phys. Rev. Lett.}
  {\bfseries 123} (2019) 241803}
  [\href{https://arxiv.org/abs/1907.12771}{{\ttfamily 1907.12771}}].

\bibitem{Baxter:2019pnz}
D.~Baxter, Y.~Kahn and G.~Krnjaic, \emph{{Electron Ionization via Dark
  Matter-Electron Scattering and the Migdal Effect}},
  \href{https://doi.org/10.1103/PhysRevD.101.076014}{\emph{Phys. Rev. D}
  {\bfseries 101} (2020) 076014}
  [\href{https://arxiv.org/abs/1908.00012}{{\ttfamily 1908.00012}}].

\bibitem{Essig:2019xkx}
R.~Essig, J.~Pradler, M.~Sholapurkar and T.-T. Yu, \emph{{Relation between the
  Migdal Effect and Dark Matter-Electron Scattering in Isolated Atoms and
  Semiconductors}},
  \href{https://doi.org/10.1103/PhysRevLett.124.021801}{\emph{Phys. Rev. Lett.}
  {\bfseries 124} (2020) 021801}
  [\href{https://arxiv.org/abs/1908.10881}{{\ttfamily 1908.10881}}].

\bibitem{Liang:2019nnx}
Z.-L. Liang, L.~Zhang, F.~Zheng and P.~Zhang, \emph{{Describing Migdal effects
  in diamond crystal with atom-centered localized Wannier functions}},
  \href{https://doi.org/10.1103/PhysRevD.102.043007}{\emph{Phys. Rev. D}
  {\bfseries 102} (2020) 043007}
  [\href{https://arxiv.org/abs/1912.13484}{{\ttfamily 1912.13484}}].

\bibitem{Kozaczuk:2020uzb}
J.~Kozaczuk and T.~Lin, \emph{{Plasmon production from dark matter
  scattering}}, \href{https://doi.org/10.1103/PhysRevD.101.123012}{\emph{Phys.
  Rev. D} {\bfseries 101} (2020) 123012}
  [\href{https://arxiv.org/abs/2003.12077}{{\ttfamily 2003.12077}}].

\bibitem{SENSEI:2020dpa}
{\scshape SENSEI} collaboration, \emph{{SENSEI: Direct-Detection Results on
  sub-GeV Dark Matter from a New Skipper-CCD}},
  \href{https://doi.org/10.1103/PhysRevLett.125.171802}{\emph{Phys. Rev. Lett.}
  {\bfseries 125} (2020) 171802}
  [\href{https://arxiv.org/abs/2004.11378}{{\ttfamily 2004.11378}}].

\bibitem{Liu:2020pat}
C.~P. Liu, C.-P. Wu, H.-C. Chi and J.-W. Chen, \emph{{Model-independent
  determination of the Migdal effect via photoabsorption}},
  \href{https://doi.org/10.1103/PhysRevD.102.121303}{\emph{Phys. Rev. D}
  {\bfseries 102} (2020) 121303}
  [\href{https://arxiv.org/abs/2007.10965}{{\ttfamily 2007.10965}}].

\bibitem{Pindzola:2020whx}
M.~S. Pindzola, J.~Colgan and M.~F. Ciappina, \emph{{Neutron ionization of
  helium near the neutron-alpha particle collision resonance}},
  \href{https://doi.org/10.1088/1361-6455/aba9af}{\emph{J. Phys. B} {\bfseries
  53} (2020) 205201}.

\bibitem{Kahn:2020fef}
Y.~Kahn, G.~Krnjaic and B.~Mandava, \emph{{Dark Matter Detection with Bound
  Nuclear Targets: The Poisson Phonon Tail}},
  \href{https://doi.org/10.1103/PhysRevLett.127.081804}{\emph{Phys. Rev. Lett.}
  {\bfseries 127} (2021) 081804}
  [\href{https://arxiv.org/abs/2011.09477}{{\ttfamily 2011.09477}}].

\bibitem{Knapen:2020aky}
S.~Knapen, J.~Kozaczuk and T.~Lin, \emph{{Migdal Effect in Semiconductors}},
  \href{https://doi.org/10.1103/PhysRevLett.127.081805}{\emph{Phys. Rev. Lett.}
  {\bfseries 127} (2021) 081805}
  [\href{https://arxiv.org/abs/2011.09496}{{\ttfamily 2011.09496}}].

\bibitem{Knapen:2021bwg}
S.~Knapen, J.~Kozaczuk and T.~Lin, \emph{{python package for dark matter
  scattering in dielectric targets}},
  \href{https://doi.org/10.1103/PhysRevD.105.015014}{\emph{Phys. Rev. D}
  {\bfseries 105} (2022) 015014}
  [\href{https://arxiv.org/abs/2104.12786}{{\ttfamily 2104.12786}}].

\bibitem{Liang:2022xbu}
Z.-L. Liang, C.~Mo, F.~Zheng and P.~Zhang, \emph{{Phonon-mediated Migdal effect
  in semiconductor detectors}},
  \href{https://doi.org/10.1103/PhysRevD.106.043004}{\emph{Phys. Rev. D}
  {\bfseries 106} (2022) 043004}
  [\href{https://arxiv.org/abs/2205.03395}{{\ttfamily 2205.03395}}].

\bibitem{LUX:2020yym}
{\scshape LUX} collaboration, \emph{{Improving sensitivity to low-mass dark
  matter in LUX using a novel electrode background mitigation technique}},
  \href{https://doi.org/10.1103/PhysRevD.104.012011}{\emph{Phys. Rev. D}
  {\bfseries 104} (2021) 012011}
  [\href{https://arxiv.org/abs/2011.09602}{{\ttfamily 2011.09602}}].

\bibitem{Liang:2020ryg}
Z.-L. Liang, C.~Mo, F.~Zheng and P.~Zhang, \emph{{Describing the Migdal effect
  with a bremsstrahlung-like process and many-body effects}},
  \href{https://doi.org/10.1103/PhysRevD.104.056009}{\emph{Phys. Rev. D}
  {\bfseries 104} (2021) 056009}
  [\href{https://arxiv.org/abs/2011.13352}{{\ttfamily 2011.13352}}].

\bibitem{He:2020sat}
H.-J. He, Y.-C. Wang and J.~Zheng, \emph{{GeV-scale inelastic dark matter with
  dark photon mediator via direct detection and cosmological and laboratory
  constraints}}, \href{https://doi.org/10.1103/PhysRevD.104.115033}{\emph{Phys.
  Rev. D} {\bfseries 104} (2021) 115033}
  [\href{https://arxiv.org/abs/2012.05891}{{\ttfamily 2012.05891}}].

\bibitem{Flambaum:2020xxo}
V.~V. Flambaum, L.~Su, L.~Wu and B.~Zhu, \emph{{New strong bounds on sub-GeV
  dark matter from boosted and Migdal effects}},
  \href{https://doi.org/10.1007/s11433-022-2090-7}{\emph{Sci. China Phys. Mech.
  Astron.} {\bfseries 66} (2023) 271011}
  [\href{https://arxiv.org/abs/2012.09751}{{\ttfamily 2012.09751}}].

\bibitem{Akerib:2021pfd}
D.~S. Akerib et~al., \emph{{Enhancing the sensitivity of the LUX-ZEPLIN (LZ)
  dark matter experiment to low energy signals}},
  \href{https://arxiv.org/abs/2101.08753}{{\ttfamily 2101.08753}}.

\bibitem{Collar:2021fcl}
J.~I. Collar, A.~R.~L. Kavner and C.~M. Lewis, \emph{{Germanium response to
  sub-keV nuclear recoils: a multipronged experimental characterization}},
  \href{https://doi.org/10.1103/PhysRevD.103.122003}{\emph{Phys. Rev. D}
  {\bfseries 103} (2021) 122003}
  [\href{https://arxiv.org/abs/2102.10089}{{\ttfamily 2102.10089}}].

\bibitem{Bell:2021zkr}
N.~F. Bell, J.~B. Dent, B.~Dutta, S.~Ghosh, J.~Kumar and J.~L. Newstead,
  \emph{{Low-mass inelastic dark matter direct detection via the Migdal
  effect}}, \href{https://doi.org/10.1103/PhysRevD.104.076013}{\emph{Phys. Rev.
  D} {\bfseries 104} (2021) 076013}
  [\href{https://arxiv.org/abs/2103.05890}{{\ttfamily 2103.05890}}].

\bibitem{Liao:2021yog}
J.~Liao, H.~Liu and D.~Marfatia, \emph{{Coherent neutrino scattering and the
  Migdal effect on the quenching factor}},
  \href{https://doi.org/10.1103/PhysRevD.104.015005}{\emph{Phys. Rev. D}
  {\bfseries 104} (2021) 015005}
  [\href{https://arxiv.org/abs/2104.01811}{{\ttfamily 2104.01811}}].

\bibitem{Menshikov:2021jxa}
L.~I. Men'shikov, P.~L. Men'shikov and M.~P. Faifman, \emph{{Relation of the
  Probability of Recoil Atom Ionization to Experimental Data on Ionization by
  Photons and Electrons}},
  \href{https://doi.org/10.1134/S1547477121020114}{\emph{Phys. Part. Nucl.
  Lett.} {\bfseries 18} (2021) 173}.

\bibitem{Acevedo:2021kly}
J.~F. Acevedo, J.~Bramante and A.~Goodman, \emph{{Accelerating composite dark
  matter discovery with nuclear recoils and the Migdal effect}},
  \href{https://doi.org/10.1103/PhysRevD.105.023012}{\emph{Phys. Rev. D}
  {\bfseries 105} (2022) 023012}
  [\href{https://arxiv.org/abs/2108.10889}{{\ttfamily 2108.10889}}].

\bibitem{COSINE-100:2021poy}
{\scshape COSINE-100} collaboration, \emph{{Searching for low-mass dark matter
  via the Migdal effect in COSINE-100}},
  \href{https://doi.org/10.1103/PhysRevD.105.042006}{\emph{Phys. Rev. D}
  {\bfseries 105} (2022) 042006}
  [\href{https://arxiv.org/abs/2110.05806}{{\ttfamily 2110.05806}}].

\bibitem{Berlin:2021zbv}
A.~Berlin, H.~Liu, M.~Pospelov and H.~Ramani, \emph{{Low-energy signals from
  the formation of dark-matter\textendash{}nucleus bound states}},
  \href{https://doi.org/10.1103/PhysRevD.105.095028}{\emph{Phys. Rev. D}
  {\bfseries 105} (2022) 095028}
  [\href{https://arxiv.org/abs/2110.06217}{{\ttfamily 2110.06217}}].

\bibitem{CDEX:2021cll}
{\scshape CDEX} collaboration, \emph{{Studies of the Earth shielding effect to
  direct dark matter searches at the China Jinping Underground Laboratory}},
  \href{https://doi.org/10.1103/PhysRevD.105.052005}{\emph{Phys. Rev. D}
  {\bfseries 105} (2022) 052005}
  [\href{https://arxiv.org/abs/2111.11243}{{\ttfamily 2111.11243}}].

\bibitem{Wang:2021oha}
W.~Wang, K.-Y. Wu, L.~Wu and B.~Zhu, \emph{{Direct detection of spin-dependent
  sub-GeV dark matter via Migdal effect}},
  \href{https://doi.org/10.1016/j.nuclphysb.2022.115907}{\emph{Nucl. Phys. B}
  {\bfseries 983} (2022) 115907}
  [\href{https://arxiv.org/abs/2112.06492}{{\ttfamily 2112.06492}}].

\bibitem{Bell:2021ihi}
N.~F. Bell, J.~B. Dent, R.~F. Lang, J.~L. Newstead and A.~C. Ritter,
  \emph{{Observing the Migdal effect from nuclear recoils of neutral particles
  with liquid xenon and argon detectors}},
  \href{https://doi.org/10.1103/PhysRevD.105.096015}{\emph{Phys. Rev. D}
  {\bfseries 105} (2022) 096015}
  [\href{https://arxiv.org/abs/2112.08514}{{\ttfamily 2112.08514}}].

\bibitem{Abbamonte:2022rfh}
P.~Abbamonte, D.~Baxter, Y.~Kahn, G.~Krnjaic, N.~Kurinsky, B.~Mandava et~al.,
  \emph{{Revisiting the dark matter interpretation of excess rates in
  semiconductors}},
  \href{https://doi.org/10.1103/PhysRevD.105.123002}{\emph{Phys. Rev. D}
  {\bfseries 105} (2022) 123002}
  [\href{https://arxiv.org/abs/2202.03436}{{\ttfamily 2202.03436}}].

\bibitem{Chatterjee:2022gbo}
S.~Chatterjee and R.~Laha, \emph{{Explorations of pseudo-Dirac dark matter
  having keV splittings and interacting via transition electric and magnetic
  dipole moments}},
  \href{https://doi.org/10.1103/PhysRevD.107.083036}{\emph{Phys. Rev. D}
  {\bfseries 107} (2023) 083036}
  [\href{https://arxiv.org/abs/2202.13339}{{\ttfamily 2202.13339}}].

\bibitem{EDELWEISS:2022ktt}
{\scshape EDELWEISS} collaboration, \emph{{Search for sub-GeV dark matter via
  the Migdal effect with an EDELWEISS germanium detector with NbSi
  transition-edge sensors}},
  \href{https://doi.org/10.1103/PhysRevD.106.062004}{\emph{Phys. Rev. D}
  {\bfseries 106} (2022) 062004}
  [\href{https://arxiv.org/abs/2203.03993}{{\ttfamily 2203.03993}}].

\bibitem{Angevaare:2022zhy}
J.~R. Angevaare, G.~Bertone, A.~P. Colijn, M.~P. Decowski and B.~J. Kavanagh,
  \emph{{Complementarity of direct detection experiments in search of light
  Dark Matter}},
  \href{https://doi.org/10.1088/1475-7516/2022/10/004}{\emph{JCAP} {\bfseries
  10} (2022) 004} [\href{https://arxiv.org/abs/2204.01580}{{\ttfamily
  2204.01580}}].

\bibitem{Campbell-Deem:2022fqm}
B.~Campbell-Deem, S.~Knapen, T.~Lin and E.~Villarama, \emph{{Dark matter direct
  detection from the single phonon to the nuclear recoil regime}},
  \href{https://doi.org/10.1103/PhysRevD.106.036019}{\emph{Phys. Rev. D}
  {\bfseries 106} (2022) 036019}
  [\href{https://arxiv.org/abs/2205.02250}{{\ttfamily 2205.02250}}].

\bibitem{DarkSide:2022dhx}
{\scshape DarkSide} collaboration, \emph{{Search for
  Dark-Matter\textendash{}Nucleon Interactions via Migdal Effect with
  DarkSide-50}},
  \href{https://doi.org/10.1103/PhysRevLett.130.101001}{\emph{Phys. Rev. Lett.}
  {\bfseries 130} (2023) 101001}
  [\href{https://arxiv.org/abs/2207.11967}{{\ttfamily 2207.11967}}].

\bibitem{Blanco:2022pkt}
C.~Blanco, I.~Harris, Y.~Kahn, B.~Lillard and J.~P\'erez-R\'\i{}os,
  \emph{{Molecular Migdal effect}},
  \href{https://doi.org/10.1103/PhysRevD.106.115015}{\emph{Phys. Rev. D}
  {\bfseries 106} (2022) 115015}
  [\href{https://arxiv.org/abs/2208.09002}{{\ttfamily 2208.09002}}].

\bibitem{Cox:2022ekg}
P.~Cox, M.~J. Dolan, C.~McCabe and H.~M. Quiney, \emph{{Precise predictions and
  new insights for atomic ionization from the Migdal effect}},
  \href{https://doi.org/10.1103/PhysRevD.107.035032}{\emph{Phys. Rev. D}
  {\bfseries 107} (2023) 035032}
  [\href{https://arxiv.org/abs/2208.12222}{{\ttfamily 2208.12222}}].

\bibitem{GlobalArgonDarkMatter:2022ppc}
{\scshape Global Argon Dark Matter} collaboration, \emph{{Sensitivity
  projections for a dual-phase argon TPC optimized for light dark matter
  searches through the ionization channel}},
  \href{https://doi.org/10.1103/PhysRevD.107.112006}{\emph{Phys. Rev. D}
  {\bfseries 107} (2023) 112006}
  [\href{https://arxiv.org/abs/2209.01177}{{\ttfamily 2209.01177}}].

\bibitem{Tomar:2022ofh}
G.~Tomar, S.~Kang and S.~Scopel, \emph{{Low-mass extension of direct detection
  bounds on WIMP-quark and WIMP-gluon effective interactions using the Migdal
  effect}},
  \href{https://doi.org/10.1016/j.astropartphys.2023.102851}{\emph{Astropart.
  Phys.} {\bfseries 150} (2023) 102851}
  [\href{https://arxiv.org/abs/2210.00199}{{\ttfamily 2210.00199}}].

\bibitem{Adams:2022zvg}
D.~Adams, D.~Baxter, H.~Day, R.~Essig and Y.~Kahn, \emph{{Measuring the Migdal
  effect in semiconductors for dark matter detection}},
  \href{https://doi.org/10.1103/PhysRevD.107.L041303}{\emph{Phys. Rev. D}
  {\bfseries 107} (2023) L041303}
  [\href{https://arxiv.org/abs/2210.04917}{{\ttfamily 2210.04917}}].

\bibitem{Berghaus:2022pbu}
K.~V. Berghaus, A.~Esposito, R.~Essig and M.~Sholapurkar, \emph{{The Migdal
  effect in semiconductors for dark matter with masses below \ensuremath{\sim}
  100 MeV}}, \href{https://doi.org/10.1007/JHEP01(2023)023}{\emph{JHEP}
  {\bfseries 01} (2023) 023}
  [\href{https://arxiv.org/abs/2210.06490}{{\ttfamily 2210.06490}}].

\bibitem{Li:2022acp}
J.~Li, L.~Su, L.~Wu and B.~Zhu, \emph{{Spin-dependent sub-GeV inelastic dark
  matter-electron scattering and Migdal effect. Part I. Velocity independent
  operator}}, \href{https://doi.org/10.1088/1475-7516/2023/04/020}{\emph{JCAP}
  {\bfseries 04} (2023) 020}
  [\href{https://arxiv.org/abs/2210.15474}{{\ttfamily 2210.15474}}].

\bibitem{NEWS-G:2023qwh}
{\scshape NEWS-G} collaboration, \emph{{Exploring light dark matter with the
  DarkSPHERE spherical proportional counter electroformed underground at the
  Boulby Underground Laboratory}},
  \href{https://doi.org/10.1103/PhysRevD.108.112006}{\emph{Phys. Rev. D}
  {\bfseries 108} (2023) 112006}
  [\href{https://arxiv.org/abs/2301.05183}{{\ttfamily 2301.05183}}].

\bibitem{DarkSide-50:2023fcw}
{\scshape DarkSide-50} collaboration, \emph{{Search for low mass dark matter in
  DarkSide-50: the bayesian network approach}},
  \href{https://doi.org/10.1140/epjc/s10052-023-11410-4}{\emph{Eur. Phys. J. C}
  {\bfseries 83} (2023) 322}
  [\href{https://arxiv.org/abs/2302.01830}{{\ttfamily 2302.01830}}].

\bibitem{SuperCDMS:2023sql}
{\scshape SuperCDMS} collaboration, \emph{{Search for low-mass dark matter via
  bremsstrahlung radiation and the Migdal effect in SuperCDMS}},
  \href{https://doi.org/10.1103/PhysRevD.107.112013}{\emph{Phys. Rev. D}
  {\bfseries 107} (2023) 112013}
  [\href{https://arxiv.org/abs/2302.09115}{{\ttfamily 2302.09115}}].

\bibitem{Bell:2023uvf}
N.~F. Bell, P.~Cox, M.~J. Dolan, J.~L. Newstead and A.~C. Ritter,
  \emph{{Exploring light dark matter with the Migdal effect in hydrogen-doped
  liquid xenon}},
  \href{https://doi.org/10.1103/PhysRevD.109.L091902}{\emph{Phys. Rev. D}
  {\bfseries 109} (2024) L091902}
  [\href{https://arxiv.org/abs/2305.04690}{{\ttfamily 2305.04690}}].

\bibitem{Tilly:2023fhw}
{\scshape MIGDAL} collaboration, \emph{{3D track reconstruction of low-energy
  electrons in the MIGDAL low pressure optical time projection chamber}},
  \href{https://doi.org/10.1088/1748-0221/18/07/C07013}{\emph{JINST} {\bfseries
  18} (2023) C07013} [\href{https://arxiv.org/abs/2307.10477}{{\ttfamily
  2307.10477}}].

\bibitem{AtzoriCorona:2023ais}
M.~Atzori~Corona, M.~Cadeddu, N.~Cargioli, F.~Dordei and C.~Giunti, \emph{{On
  the impact of the Migdal effect in reactor CE\ensuremath{\nu}NS
  experiments}},
  \href{https://doi.org/10.1016/j.physletb.2024.138627}{\emph{Phys. Lett. B}
  {\bfseries 852} (2024) 138627}
  [\href{https://arxiv.org/abs/2307.12911}{{\ttfamily 2307.12911}}].

\bibitem{Qiao:2023pbw}
M.~Qiao, C.~Xia and Y.-F. Zhou, \emph{{Diurnal modulation of electron recoils
  from DM-nucleon scattering through the Migdal effect}},
  \href{https://doi.org/10.1088/1475-7516/2023/11/079}{\emph{JCAP} {\bfseries
  11} (2023) 079} [\href{https://arxiv.org/abs/2307.12820}{{\ttfamily
  2307.12820}}].

\bibitem{Xu:2023wev}
J.~Xu et~al., \emph{{Search for the Migdal effect in liquid xenon with
  keV-level nuclear recoils}},
  \href{https://doi.org/10.1103/PhysRevD.109.L051101}{\emph{Phys. Rev. D}
  {\bfseries 109} (2024) L051101}
  [\href{https://arxiv.org/abs/2307.12952}{{\ttfamily 2307.12952}}].

\bibitem{LZ:2023poo}
{\scshape LZ} collaboration, \emph{{Search for new physics in low-energy
  electron recoils from the first LZ exposure}},
  \href{https://doi.org/10.1103/PhysRevD.108.072006}{\emph{Phys. Rev. D}
  {\bfseries 108} (2023) 072006}
  [\href{https://arxiv.org/abs/2307.15753}{{\ttfamily 2307.15753}}].

\bibitem{PandaX:2023xgl}
{\scshape PandaX} collaboration, \emph{{Search for
  Dark-Matter\textendash{}Nucleon Interactions with a Dark Mediator in
  PandaX-4T}},
  \href{https://doi.org/10.1103/PhysRevLett.131.191002}{\emph{Phys. Rev. Lett.}
  {\bfseries 131} (2023) 191002}
  [\href{https://arxiv.org/abs/2308.01540}{{\ttfamily 2308.01540}}].

\bibitem{Yun:2023huf}
Z.~Yun, J.~Sun, B.~Zhu and X.~Liu, \emph{{Probing inelastic signatures of dark
  matter detection via polarized nucleus*}},
  \href{https://doi.org/10.1088/1674-1137/ad6416}{\emph{Chin. Phys. C}
  {\bfseries 48} (2024) 103106}
  [\href{https://arxiv.org/abs/2309.01203}{{\ttfamily 2309.01203}}].

\bibitem{Gu:2023pfg}
Y.~Gu, J.~Tang, L.~Wu and B.~Zhu, \emph{{Probing light DM via the Migdal effect
  with spherical proportional counter*}},
  \href{https://doi.org/10.1088/1674-1137/acfaef}{\emph{Chin. Phys. C}
  {\bfseries 47} (2023) 125105}
  [\href{https://arxiv.org/abs/2309.09740}{{\ttfamily 2309.09740}}].

\bibitem{Li:2023xkf}
Y.-F. Li and S.-y. Xia, \emph{{Migdal effect of phonon-mediated neutrino
  nucleus scattering in semiconductor detectors}},
  \href{https://doi.org/10.1016/j.nuclphysb.2024.116632}{\emph{Nucl. Phys. B}
  {\bfseries 1006} (2024) 116632}
  [\href{https://arxiv.org/abs/2310.05704}{{\ttfamily 2310.05704}}].

\bibitem{Herrera:2023xun}
G.~Herrera, \emph{{A neutrino floor for the Migdal effect}},
  \href{https://doi.org/10.1007/JHEP05(2024)288}{\emph{JHEP} {\bfseries 05}
  (2024) 288} [\href{https://arxiv.org/abs/2311.17719}{{\ttfamily
  2311.17719}}].

\bibitem{Kim:2023lkb}
W.~K. Kim, H.~Y. Lee, K.~W. Kim, Y.~J. Ko, J.~A. Jeon, H.~J. Kim et~al.,
  \emph{{Scintillation characteristics of an undoped CsI crystal at
  low-temperature for dark matter search}},
  \href{https://arxiv.org/abs/2312.07957}{{\ttfamily 2312.07957}}.

\bibitem{Dzuba:2023zcq}
V.~A. Dzuba, V.~V. Flambaum and I.~B. Samsonov, \emph{{Migdal-type effect in
  the dark matter absorption process}},
  \href{https://doi.org/10.1103/PhysRevD.109.115032}{\emph{Phys. Rev. D}
  {\bfseries 109} (2024) 115032}
  [\href{https://arxiv.org/abs/2312.10566}{{\ttfamily 2312.10566}}].

\bibitem{SENSEI:2023zdf}
{\scshape SENSEI} collaboration, \emph{{SENSEI: First Direct-Detection Results
  on sub-GeV Dark Matter from SENSEI at SNOLAB}},
  \href{https://arxiv.org/abs/2312.13342}{{\ttfamily 2312.13342}}.

\bibitem{Liang:2024tef}
J.-H. Liang, Y.~Liao, X.-D. Ma and H.-L. Wang, \emph{{Comprehensive constraints
  on fermionic dark matter-quark tensor interactions in direct detection
  experiments}},  \href{https://arxiv.org/abs/2401.05005}{{\ttfamily
  2401.05005}}.

\bibitem{He:2024hkr}
H.-J. He, Y.-C. Wang and J.~Zheng, \emph{{Probing light inelastic dark matter
  from direct detection}},
  \href{https://doi.org/10.1016/j.dark.2024.101670}{\emph{Phys. Dark Univ.}
  {\bfseries 46} (2024) 101670}
  [\href{https://arxiv.org/abs/2403.03128}{{\ttfamily 2403.03128}}].

\bibitem{Balan:2024cmq}
S.~Balan et~al., \emph{{Resonant or asymmetric: The status of sub-GeV dark
  matter}},  \href{https://arxiv.org/abs/2405.17548}{{\ttfamily 2405.17548}}.

\bibitem{MIGDAL:2024alc}
{\scshape MIGDAL} collaboration, \emph{{Transforming a rare event search into a
  not-so-rare event search in real-time with deep learning-based object
  detection}},  \href{https://arxiv.org/abs/2406.07538}{{\ttfamily
  2406.07538}}.

\bibitem{Kang:2024kec}
S.~Kang, S.~Scopel and G.~Tomar, \emph{{Low-mass constraints on WIMP effective
  models of inelastic scattering using the Migdal effect}},
  \href{https://arxiv.org/abs/2407.16187}{{\ttfamily 2407.16187}}.

\bibitem{Ibarra:2024mpq}
A.~Ibarra, M.~Reichard and G.~Tomar, \emph{{Probing Dark Matter Electromagnetic
  Properties in Direct Detection Experiments}},
  \href{https://arxiv.org/abs/2408.15760}{{\ttfamily 2408.15760}}.

\bibitem{Lee:2024wzd}
D.~H. Lee et~al., \emph{{COSINE-100U: Upgrading the COSINE-100 Experiment for
  Enhanced Sensitivity to Low-Mass Dark Matter Detection}},
  \href{https://arxiv.org/abs/2409.15748}{{\ttfamily 2409.15748}}.

\bibitem{Kouvaris:2016afs}
C.~Kouvaris and J.~Pradler, \emph{{Probing sub-GeV Dark Matter with
  conventional detectors}},
  \href{https://doi.org/10.1103/PhysRevLett.118.031803}{\emph{Phys. Rev. Lett.}
  {\bfseries 118} (2017) 031803}
  [\href{https://arxiv.org/abs/1607.01789}{{\ttfamily 1607.01789}}].

\bibitem{Millar:2018hkv}
A.~Millar, G.~Raffelt, L.~Stodolsky and E.~Vitagliano, \emph{{Neutrino mass
  from bremsstrahlung endpoint in coherent scattering on nuclei}},
  \href{https://doi.org/10.1103/PhysRevD.98.123006}{\emph{Phys. Rev. D}
  {\bfseries 98} (2018) 123006}
  [\href{https://arxiv.org/abs/1810.06584}{{\ttfamily 1810.06584}}].

\bibitem{johnson2007atomic}
W.~R. Johnson, \emph{Atomic structure theory}. Springer, 2007.

\bibitem{sakurai1967advanced}
J.~J. Sakurai, \emph{Advanced quantum mechanics}. Pearson Education India,
  1967.

\bibitem{peixoto1969elastic}
E.~Peixoto, C.~Bunge and R.~Bonham, \emph{Elastic and inelastic electron
  scattering by he and ne atoms in their ground states}, {\emph{Physical
  Review} {\bfseries 181} (1969) 322}.

\bibitem{naon1972importance}
M.~Naon and M.~Cornille, \emph{The importance of electronic correlations for
  the calculation of x-ray and fast electron scattering by ne (1s)},
  {\emph{Journal of Physics B: Atomic and Molecular Physics} {\bfseries 5}
  (1972) 1965}.

\bibitem{khatun2021theoretical}
M.~M. Khatun, M.~Haque, M.~A.~R. Patoary, M.~Shorifuddoza, M.~H. Khandker,
  A.~F. Haque et~al., \emph{Theoretical study of e$\pm$scattering by the au
  atom}, {\emph{Results in Physics} {\bfseries 29} (2021) 104742}.

\bibitem{parvin2023theoretical}
S.~Parvin, M.~M. Billah, M.~H. Khandker, M.~I. Hossain, M.~Haque,
  M.~Shahmohammadi~Beni et~al., \emph{A theoretical study of scattering of
  e$\pm$by tl atom}, {\emph{Atoms} {\bfseries 11} (2023) 37}.

\bibitem{Kahn:2021ttr}
Y.~Kahn and T.~Lin, \emph{{Searches for light dark matter using condensed
  matter systems}}, \href{https://doi.org/10.1088/1361-6633/ac5f63}{\emph{Rept.
  Prog. Phys.} {\bfseries 85} (2022) 066901}
  [\href{https://arxiv.org/abs/2108.03239}{{\ttfamily 2108.03239}}].

\bibitem{gozem2015photoelectron}
S.~Gozem, A.~O. Gunina, T.~Ichino, D.~L. Osborn, J.~F. Stanton and A.~I.
  Krylov, \emph{Photoelectron wave function in photoionization: Plane wave or
  coulomb wave?}, {\emph{The journal of physical chemistry letters} {\bfseries
  6} (2015) 4532}.

\bibitem{doi:10.1139/p07-197}
M.~F. Gu, \emph{The flexible atomic code},
  \href{https://doi.org/10.1139/p07-197}{\emph{Canadian Journal of Physics}
  {\bfseries 86} (2008) 675}
  [\href{https://arxiv.org/abs/https://doi.org/10.1139/p07-197}{{\ttfamily
  https://doi.org/10.1139/p07-197}}].

\bibitem{Phelps:2024}
\emph{{Phelps Database, retrieved Sep 25, 2024}},
  \href{https://arxiv.org/abs/http://www.lxcat.laplace.univ-tlse.fr/}{{\ttfamily
  http://www.lxcat.laplace.univ-tlse.fr/}}.

\bibitem{Morgan:2024}
\emph{{Morgan Database, retrieved Oct 02, 2024}},
  \href{https://arxiv.org/abs/http://www.lxcat.laplace.univ-tlse.fr/}{{\ttfamily
  http://www.lxcat.laplace.univ-tlse.fr/}}.

\bibitem{bonham1974high}
R.~Bonham and M.~Fink, \emph{High-energy Electron Scattering}, ACS monograph.
  Van Nostrand Reinhold, 1974.

\bibitem{garcia1999electron}
G.~Garcia, M.~Roteta, F.~Manero, F.~Blanco and A.~Williart, \emph{Electron
  scattering by ne, ar and kr at intermediate and high energies, 0.5-10 kev},
  {\emph{Journal of Physics B: Atomic, Molecular and Optical Physics}
  {\bfseries 32} (1999) 1783}.

\bibitem{inokuti1971inelastic}
M.~Inokuti, \emph{Inelastic collisions of fast charged particles with atoms and
  molecules―the bethe theory revisited}, {\emph{Reviews of modern physics}
  {\bfseries 43} (1971) 297}.

\bibitem{inokuti1975total}
M.~Inokuti, R.~P. Saxon and J.~Dehmer, \emph{Total cross-sections for inelastic
  scattering of charged particles by atoms and molecules―viii. systematics for
  atoms in the first and second row}, {\emph{International Journal for
  Radiation Physics and Chemistry} {\bfseries 7} (1975) 109}.

\bibitem{inokuti1981oscillator}
M.~Inokuti, J.~L. Dehmer, T.~Baer and J.~Hanson, \emph{Oscillator-strength
  moments, stopping powers, and total inelastic-scattering cross sections of
  all atoms through strontium}, {\emph{Physical Review A} {\bfseries 23} (1981)
  95}.

\bibitem{salvat2022inelastic}
F.~Salvat, L.~Barjuan and P.~Andreo, \emph{Inelastic collisions of fast charged
  particles with atoms: Bethe asymptotic formulas and shell corrections},
  {\emph{Physical Review A} {\bfseries 105} (2022) 042813}.

\bibitem{atrazhev1981hot}
V.~Atrazhev and I.~Iakubov, \emph{Hot electrons in non-polar liquids},
  {\emph{Journal of Physics C: Solid State Physics} {\bfseries 14} (1981)
  5139}.

\bibitem{atrazhev1985heating}
V.~Atrazhev and E.~Dmitriev, \emph{Heating and diffusion of hot electrons in
  non-polar liquids}, {\emph{Journal of Physics C: Solid State Physics}
  {\bfseries 18} (1985) 1205}.

\end{thebibliography}\endgroup
\end{document}